Letter

# Probing the hubble tension with an infinite-future condition on the hubble parameter

Yuan-Bo Xie [a,b,1], Cheqiu Lyu [b,c,d,e,1], Zhiyuan Mo [a,b,1], Wei Hong [a,b], Tong-Jie Zhang [a,b,*]

[a] Institute for Frontiers in Astronomy and Astrophysics, Beijing Normal University, Beijing, 102206, China
[b] School of Physics and Astronomy, Beijing Normal University, Beijing, 100875, China
[c] Department of Astronomy, University of Science and Technology of China, Hefei, 230026, China
[d] Kavli Institute for Astronomy and Astrophysics, Peking University, Beijing, 100871, China
[e] Department of Astronomy, School of Physics, Peking University, Beijing, 100871, China

ABSTRACT

We study the impact of imposing an infinite-future condition on Gaussian-process (GP) reconstructions of $H(z)$ from 37 cosmic-chronometer measurements. Implementing the asymptotic limit $H(-1) = 0$ expected in constant-$w$CDM with $w > -1$ as a pseudo-point at $z = -1$ with tunable uncertainty $\sigma_{-1}$ lowers the GP-inferred Hubble constant from $H_0 = 68.71 \pm 6.08$ to $H_0 \simeq (64.67\text{–}65.86) \pm (4.45\text{–}4.85)$ km s$^{-1}$ Mpc$^{-1}$. The resulting $H_0$ remains within $\sim 0.32\text{–}0.61\,\sigma$ of the *Planck* $\Lambda$CDM value, while the separation from representative local distance-ladder measurements increases to $\sim 1.45\text{–}1.83\,\sigma$. A scan over $\sigma_{-1}$ shows that the shift is governed by the effective pseudo-point weight, interpolating between the hard-condition and data-dominated limits. Finally, constant-$w$CDM Markov Chain Monte Carlo (MCMC) fits to the same $H(z)$ data with $H_0$ fixed to the GP-inferred values show that $\Omega_m$ is only weakly affected, whereas lower $H_0$ shifts $w$ toward less negative values; allowing curvature broadens constraints and remains consistent with $\Omega_k = 0$.

## 1. Introduction

The Hubble constant ($H_0$) sets the present-day expansion rate of the Universe and is a key parameter in cosmology. In recent years, a persistent discrepancy has been reported between late-time distance-ladder measurements and early-Universe inferences from the Cosmic Microwave Background (CMB) within the base-$\Lambda$CDM model. For example, the Cepheid-calibrated Type Ia supernova (SN Ia) distance ladder finds $H_0 = 73.04 \pm 1.04$ km s$^{-1}$ Mpc$^{-1}$ [1], whereas *Planck* CMB data imply $H_0 = 67.4 \pm 0.5$ km s$^{-1}$ Mpc$^{-1}$ under base-$\Lambda$CDM [2]. The difference corresponds to an approximately $5\sigma$ tension (for two Gaussian constraints we define $n \equiv |\mu_1 - \mu_2|/\sqrt{\sigma_1^2 + \sigma_2^2}$). Despite concerted efforts to mitigate it through improved methodologies and independent cross-checks [3–6], the origin of the tension remains unsettled, motivating independent approaches to determine $H_0$ and to test late-time expansion measurements.

One of the simplest ways to obtain the value of $H_0$ is by using observational $H(z)$ data (OHD). In particular, OHD from the cosmic chronometers (CC) method have been widely used for this purpose [7–14]. The CC approach provides a direct estimate of the expansion rate via differential aging of passively evolving galaxies, using $H(z) = -(1+z)^{-1}\,dz/dt$. To reduce dependence on specific parametric cosmological models, Busti et al. [15] introduced a non-parametric reconstruction based on Gaussian Processes (GP), where the covariance function encodes correlations in the reconstructed function. Using 19 OHD points from the CC method and baryon acoustic oscillation (BAO) data, they reconstructed $H(z)$ and extrapolated to redshift zero, obtaining $H_0 = 64.9 \pm 4.2$ km s$^{-1}$ Mpc$^{-1}$ [16]. Two methodological aspects deserve attention. First, although GP reconstructions are non-parametric, the inferred results can depend on the choice of kernel, the treatment of hyperparameters, and the reconstruction implementation [17–23]. These choices can affect the reconstructed $H(z)$ and the extrapolated value of $H_0$, motivating systematic checks and robustness tests [17,24–26]. Second, in this work we restrict the OHD sample to CC measurements and do not include BAO-derived points. BAO observables are typically reported relative to a fiducial cosmology (and often involve assumptions about the sound horizon), so treating them as direct inputs to a model-agnostic GP reconstruction requires additional assumptions beyond the scope of this paper [27,28].

* Corresponding author.
*E-mail address:* tjzhang@bnu.edu.cn (T.-J. Zhang).
[1] These authors contributed equally to this work.

However, given the sizable uncertainties in current $H(z)$ measurements and their limited redshift coverage, a one-sided extrapolation to $z = 0$ based only on the available $z > 0$ data can be poorly constrained. Motivated by this, we examine how a physically motivated infinite-future condition restricts the space of admissible $H(z)$ histories and thereby affects the inferred $H_0$. We implement the infinite-future limit as a pseudo-point with a tunable uncertainty, which allows us to explicitly control its effective weight relative to the OHD and to quantify its impact on $H_0$ and other cosmological parameters. In this paper, we propose for the first time to implement the infinite-future limit as a prior pseudo-point by adopting the condition $H(z = -1) = 0$ for dark-energy equations of state $w > -1$. Although this limit is theoretically fixed within the assumed model class (i.e., it has no observational uncertainty), we still treat it as a pseudo-point with an assigned uncertainty to control its statistical weight relative to the OHD. This setup allows us to quantify how strongly such a boundary condition propagates into the inferred $H_0$ and other cosmological parameters, while making the associated model dependence explicit. In Section 2, we introduce the theoretical future condition on the Hubble parameter and the corresponding GP reconstruction framework. In Section 3, we present the main reconstruction results and their implications, including the dependence on the pseudo-point weight and the associated GP hyperparameter behavior. In Section 4, we further test the method with mock OHD realizations, while additional kernel and mean-function robustness checks are collected in Appendix A. Finally, in Section 5, we summarize the main conclusions and discuss the limitations and possible extensions of this approach.

## 2. Methodology

The Hubble parameter quantifies the cosmic expansion rate, defined as $H(t) \equiv \dot{a}/a$, where $a(t)$ is the scale factor, and the Hubble constant is $H_0 \equiv H(t_0)$. In the constant-$w$ dark-energy model ($w$CDM), the expansion history can be written as $H(z) = H_0 E(z)$ with

$$E^2(z) = \Omega_M (1+z)^3 + \Omega_k (1+z)^2 + \Omega_{\rm DE} (1+z)^{3(1+w)}, \tag{1}$$

where $\Omega_M$, $\Omega_k$, and $\Omega_{\rm DE}$ denote the present-day density parameters of matter, curvature, and dark energy, respectively, and $w$ is the (constant) equation-of-state parameter. Throughout this work we neglect the radiation contribution at late times, since $\Omega_R$ is negligible for the redshift range considered here. We consider both the spatially flat case ($\Omega_k = 0$) and a non-flat extension, imposing the closure relation $\Omega_M + \Omega_{\rm DE} + \Omega_k = 1$ (neglecting radiation).

The relation $1 + z = a_0/a(t)$ implies that, in a formal sense, $z < 0$ corresponds to future emission ($a > a_0$), while the asymptotic future limit $a \to \infty$ maps to $z \to -1$. Therefore,

$$H(z \to -1) = \begin{cases} \infty & w < -1 \ \text{(phantom)} \\ H_0 \sqrt{\Omega_{\rm DE}} & w = -1 \ (\Lambda\text{CDM}) \\ 0 & w > -1 \ \text{(quintessence-like)}. \end{cases} \tag{2}$$

Fig. 1 illustrates the late-time behavior of $H(z)$ in constant-$w$CDM models. For the quintessence-like regime $w > -1$, the expansion rate approaches a universal asymptotic limit, $\lim_{z\to-1} H(z) = 0$, which is independent of the specific choices of $(\Omega_{\rm M}, \Omega_{\rm DE}, \Omega_k)$. In this work we do not consider $w \le -1$: for $w = -1$ ($\Lambda$CDM) one has $\lim_{z\to-1} H(z) = H_0\sqrt{\Omega_\Lambda}$, while for $w < -1$ (phantom) $H(z)$ diverges as $z \to -1$.

Theoretically, the redshift is defined observationally by $1 + z \equiv \lambda_{\rm obs}/\lambda_{\rm em}$, and in a Friedmann–Lemaître–Robertson–Walker (FLRW) universe one has $1 + z = a_0/a(t_{\rm em})$. In a formal sense, negative redshift ($z < 0$) corresponds to $a(t_{\rm em}) > a_0$, i.e. the future evolution relative to the present epoch, which is not directly accessible to observations today. Nonetheless, the asymptotic limit $z \to -1$ provides a well-defined theoretical value in the infinite future, and we can still explore its impact

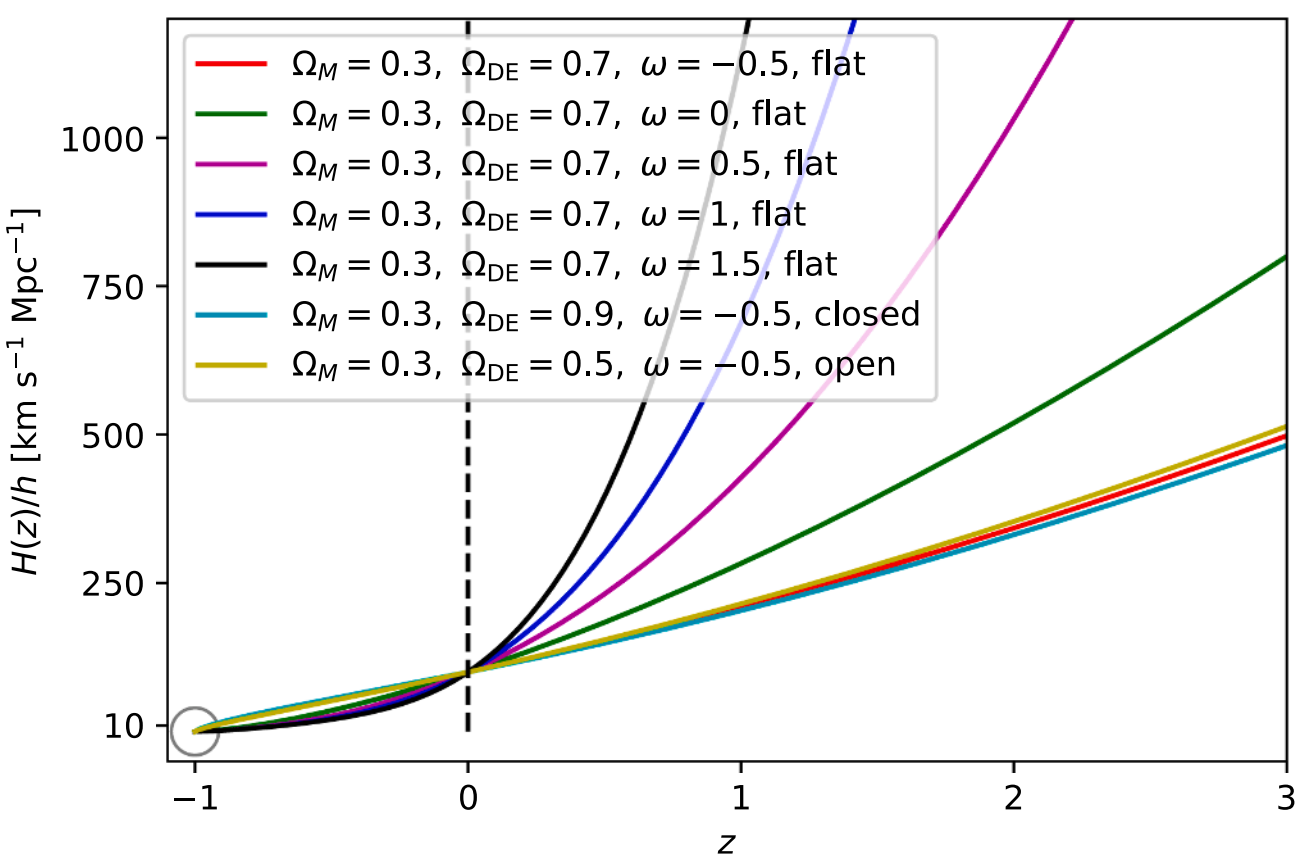

**Fig. 1.** Hubble parameter as a function of redshift for several representative constant-$w$CDM models, shown as $H(z)/h$. For $w > -1$, the curves approach the circled point $(-1, 0)$, i.e. $\lim_{z\to-1} H(z) = 0$.

[29]. Motivated by this, we impose $H(z = -1) = 0$ (denoted as $H(-1)$) as an external prior pseudo-point to constrain $H_0$.

In this paper, we use OHD derived from CC only. The dataset consists of 37 $H(z)$ measurements spanning the redshift range $z \in [0.07, 1.965]$, as listed in Table 1. The $z = 0.09$ entry originates from an $H_0$ estimate in Ref. [7], not a direct CC determination of $H(z)$. Following the correction procedure of Ref. [30], we convert it to an effective $H(0.09) = 70.7 \pm 12.3\ \ {\rm km\,s^{-1}\,Mpc^{-1}}$. We perform the Gaussian-process reconstruction over $z \in [-1, 3]$: the upper bound extends beyond the maximum data redshift, while the extension to $z < 0$ is introduced solely to implement the theoretical boundary condition at $z = -1$ (denoted as $H(-1)$). We then infer $H_0 \equiv H(z = 0)$ from the reconstructed $H(z)$.

In comparison to the OHD, we impose the infinite-future condition $H(z = -1) = 0$, which we denote as $H(-1)$. In the constant-$w$CDM framework with $w > -1$, this limit is fixed within this model class. In the GP analysis, we implement $H(-1)$ as a pseudo-point at $z = -1$ and assign it an uncertainty $\sigma_{-1}$, which sets its weight relative to the OHD. Because the GP reconstruction is performed directly on $H(z)$, this pseudo-point should be interpreted as a theoretical prior implemented directly at the level of $H(z)$ rather than as a condition enforced through the Raychaudhuri and continuity equations for the underlying energy components. At the GP level, the pseudo-point encodes only the boundary value; it does not encode the full dynamics or distinguish between different models that share the same asymptotic value.

We consider three representative choices for $\sigma_{-1}$ to illustrate and compare how different effective weights assigned to the future pseudo-point affect the GP reconstruction and the resulting parameter constraints. In the no-error (NE) case, we impose a hard condition with $\sigma_{\rm NE}[H(-1)] = 0$ (implemented numerically by an infinitesimal uncertainty). In the average-error (AE) case, we set $\sigma_{\rm AE}[H(-1)] \equiv \bar{\sigma}$, where $\bar{\sigma}$ is the mean uncertainty of the OHD sample; for our dataset, $\bar{\sigma} = 22.5373$. In the linear-error (LE) case, we model the redshift dependence of the OHD uncertainties by a simple fit $\sigma(z) = k|z| + b$, and evaluate it at $z = -1$, giving $\sigma_{\rm LE}[H(-1)] = 26.4549$. More generally, we scan the pseudo-point uncertainty $\sigma_{-1}$ over a wide range to continuously vary its weight relative to the OHD. In practice, we take $\sigma_{-1} \in [10^{-5}, 10^3]$ and sample 45 logarithmically spaced values (see Section 3). For each $\sigma_{-1}$, we add the pseudo-point $H(-1) = 0 \pm \sigma_{-1}$ to the OHD data set and perform a GP reconstruction of $H(z)$ (Matérn kernel with $\nu = 3/2$), from which we infer $H_0 \equiv H(z = 0)$.

Gaussian processes (GP) provide a non-parametric framework to reconstruct a continuous function from discrete measurements. Given training data $\{z_i, H_i\}$ with covariance matrix $\mathbf{C}$ (assumed diagonal in this work), the GP prediction at a set of test redshifts $\{z_j^*\}$ follows a multivariate Gaussian distribution. The predictive mean and covariance are fully determined by the kernel $k(z,z')$ and by $\mathbf{C}$. In this work we adopt a Matérn kernel with $\nu = 3/2$,

$$k(z,z') = \sigma_f^2 \, \frac{2^{1-\nu}}{\Gamma(\nu)} \left( \frac{\sqrt{2\nu}\,|z-z'|}{\ell} \right)^{\nu} K_\nu\!\left( \frac{\sqrt{2\nu}\,|z-z'|}{\ell} \right), \tag{3}$$

where $\sigma_f$ and $\ell$ are hyperparameters that control the typical amplitude and correlation length of the reconstructed function, respectively; $K_\nu(\cdot)$ is the modified Bessel function of the second kind and $\Gamma(\cdot)$ is the Gamma function. Following Ref. [36], we adopt this kernel choice for the OHD. We implement the GP reconstruction using the public package `scikit-learn` [37]. The hyperparameters $\theta = (\sigma_f, \ell)$ are determined by maximizing the GP log-marginal likelihood [17],

$$\ln p(\mathbf{H} \mid \mathbf{z}, \theta) = -\frac{1}{2}(\mathbf{H}-\mathbf{m})^{\mathrm{T}} \mathbf{K}_y^{-1} (\mathbf{H}-\mathbf{m}) - \frac{1}{2} \ln \left| \mathbf{K}_y \right| - \frac{n}{2} \ln(2\pi), \tag{4}$$

where $\mathbf{H} = (H_1, \dots, H_n)^{\mathrm{T}}$, $\mathbf{m}_i = m(z_i)$, and $\mathbf{K}_y = \mathbf{K}_\theta + \mathbf{C}$ with $(\mathbf{K}_\theta)_{ij} = k_\theta(z_i, z_j)$. The resulting best-fit hyperparameters are then used to obtain the reconstructed $H(z)$ and its uncertainty. Additional robustness tests of the GP inference, including alternative kernel choices and mean-function prescriptions, are presented in Appendix A.

**Table 1**
List of the 37 cosmic-chronometer $H(z)$ measurements used in this work.

| $z$ | $H(z)$ [km s$^{-1}$ Mpc$^{-1}$] | $\sigma_H$ | Ref. |
|---|---|---|---|
| 0.07 | 69.0 | 19.6 | [11] |
| 0.09 | 70.7 | 12.3 | [7] |
| 0.12 | 68.6 | 26.2 | [11] |
| 0.17 | 83.0 | 8.0 | [8] |
| 0.179 | 75.0 | 4.0 | [31] |
| 0.199 | 75.0 | 5.0 | [31] |
| 0.20 | 72.9 | 29.6 | [11] |
| 0.27 | 77.0 | 14.0 | [8] |
| 0.28 | 88.8 | 36.6 | [11] |
| 0.352 | 83.0 | 14.0 | [31] |
| 0.38 | 83.0 | 13.5 | [13] |
| 0.40 | 95.0 | 17.0 | [8] |
| 0.4004 | 77.0 | 10.2 | [13] |
| 0.425 | 87.1 | 11.2 | [13] |
| 0.4497 | 92.8 | 12.9 | [13] |
| 0.46 | 88.48 | 12.33 | [32] |
| 0.47 | 89.0 | 49.6 | [14] |
| 0.4783 | 80.9 | 9.0 | [13] |
| 0.48 | 97.0 | 62.0 | [9] |
| 0.50 | 72.1 | 34.68 | [33] |
| 0.593 | 104.0 | 13.0 | [31] |
| 0.67 | 119.45 | 17.82 | [32] |
| 0.68 | 92.0 | 8.0 | [31] |
| 0.781 | 105.0 | 12.0 | [31] |
| 0.80 | 113.1 | 25.22 | [34] |
| 0.83 | 108.28 | 18.13 | [32] |
| 0.875 | 125.0 | 17.0 | [31] |
| 0.88 | 90.0 | 40.0 | [9] |
| 0.90 | 117.0 | 23.0 | [8] |
| 1.037 | 154.0 | 20.0 | [31] |
| 1.26 | 135.0 | 65.0 | [35] |
| 1.30 | 168.0 | 17.0 | [8] |
| 1.363 | 160.0 | 33.6 | [12] |
| 1.43 | 177.0 | 18.0 | [8] |
| 1.53 | 140.0 | 14.0 | [8] |
| 1.75 | 202.0 | 40.0 | [8] |
| 1.965 | 186.5 | 50.4 | [12] |

Finally, we propagate the GP-inferred $H_0$ into parametric constant-$w$CDM fits using Markov Chain Monte Carlo (MCMC) sampling to quantify the impact of the $H(-1)$ prior (and its weight) on $H_0$ and other cosmological parameters; details and results are presented in Section 3.

## 3. Results

We first perform a GP reconstruction using the CC-only OHD sample described in Section 2. The reconstruction is shown in Fig. 2(a),

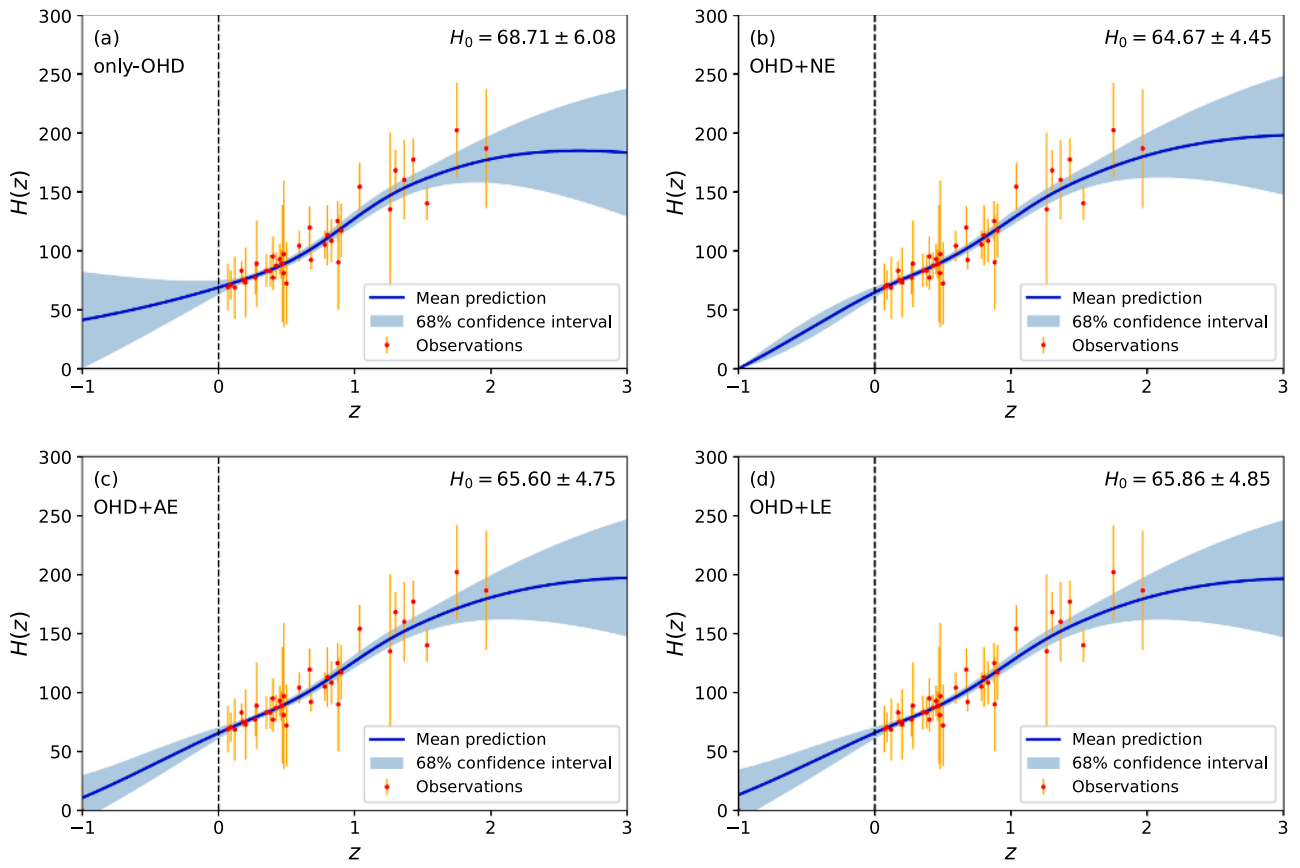

**Fig. 2.** GP reconstructions of $H(z)$ using the 37-point CC OHD sample. Panel (a) shows the OHD-only reconstruction. Panels (b)–(d) show reconstructions after adding the pseudo-point $H(-1) = 0$ at $z = -1$ under three prescriptions for its uncertainty $\sigma_{-1}$: no-error (NE), average-error (AE), and linear-error (LE), respectively. In each panel, points with error bars are the input data (including the pseudo-point when applicable), the solid curve is the GP mean, and the shaded band denotes the 68.27% ($1\sigma$) confidence region.

from which we obtain $H_0 = 68.71 \pm 6.08$ km s$^{-1}$ Mpc$^{-1}$ by evaluating the GP posterior at $z = 0$. Given the current GP uncertainty, this value is compatible with both early- and late-Universe determinations. We then repeat the reconstruction after adding the pseudo-point $H(-1) = 0$ at $z = -1$, adopting three representative choices for its uncertainty $\sigma_{-1}$ (NE/AE/LE) as defined in Section 2. The corresponding results are presented in Fig. 2(b)–(d). In the hard-limit case (NE) we obtain $H_0 = 64.67 \pm 4.45$ km s$^{-1}$ Mpc$^{-1}$, which is slightly lower than the AE and LE cases, $H_0 = 65.60 \pm 4.75$ and $H_0 = 65.86 \pm 4.85$ km s$^{-1}$ Mpc$^{-1}$, respectively. Overall, introducing the $H(-1)$ pseudo-point shifts the inferred $H_0$ to lower values compared to the OHD-only reconstruction, with the magnitude of the shift controlled by the effective weight (i.e., $\sigma_{-1}$) assigned to the pseudo-point.

These results show that incorporating the pseudo-point $H(-1) = 0$ shifts the GP-inferred $H_0$ to lower values relative to the OHD-only reconstruction. Quantitatively, the shifts correspond to a $0.37$–$0.54\,\sigma$ difference with respect to the OHD-only result. For the AE and LE prescriptions, the pseudo-point is assigned a finite uncertainty, so the reconstruction is only weakly constrained near $z = -1$; consequently, the posterior band broadens toward $z = -1$ and is substantially larger there than around $z = 0$. Although the statistical significance of the shift remains modest given the current GP uncertainties, adding the $H(-1)$ prior systematically lowers the mean inferred $H_0$, thereby increasing the separation from representative local distance-ladder determinations. Numerically, the resulting $H_0$ values are closer to early-Universe benchmark $H_0$ inferences, such as the *Planck* base-$\Lambda$CDM inference. We emphasize, however, that this is only a parameter-level comparison in $H_0$: the imposed future condition $H(-1) = 0$ is motivated within constant-$w$CDM with $w > -1$ and does not correspond to the asymptotic $z \to -1$ behavior of $H(z)$ in $\Lambda$CDM. For reference, the Dark Energy Spectroscopic Instrument (DESI) Data Release 2 (DR2) BAO calibration using Big Bang nucleosynthesis (BBN) yields $H_0 = 68.51 \pm 0.58$ km s$^{-1}$ Mpc$^{-1}$ [38]. In terms of the discrepancy metric introduced in Section 1, the NE/AE/LE cases differ from the *Planck* and DESI + BBN values by only $\sim 0.32$–$0.61\,\sigma$ and $\sim 0.54$–$0.86\,\sigma$, respectively, whereas the offset relative to the distance ladder increases to $\sim 1.45$–$1.83\,\sigma$. A comparison of these $H_0$ constraints is shown in Fig. 3. Given that the current GP uncertainties remain dominant, we refrain from further physical interpretation and focus on quantifying the dependence on the adopted pseudo-point weight.

Although $H(-1) = 0$ is an exact boundary condition within the considered model class (with NE representing the strong-limit), we scan $\sigma_{-1}$ to quantify how sensitively the inferred $H_0$ depends on how strongly this condition is enforced; AE/LE are simply representative choices in the transition regime. Using the same GP setup as above (Matérn kernel with $\nu = 3/2$) and the OHD sample including the pseudo-point $H(-1) = 0$ at $z = -1$, we repeat the reconstruction while scanning the pseudo-point uncertainty $\sigma_{-1}$. Specifically, we sample $\sigma_{-1} \in [10^{-5}, 10^3]$ with 45 logarithmically spaced values, and for each value we extract $H_0$ from the reconstructed GP mean at $z = 0$. To highlight the systematic shift, we plot $\Delta H_0 \equiv H_0(\sigma_{-1}) - H_0^{\text{only-OHD}}$. In the same scan we also track the corresponding best-fit GP hyperparameters, namely $\sigma_f$ and $\ell$, as well as the normalized kernel correlation between two redshifts [17],

$$\rho(z, z') \equiv \frac{k(z, z')}{\sqrt{k(z, z)\, k(z', z')}}. \tag{5}$$

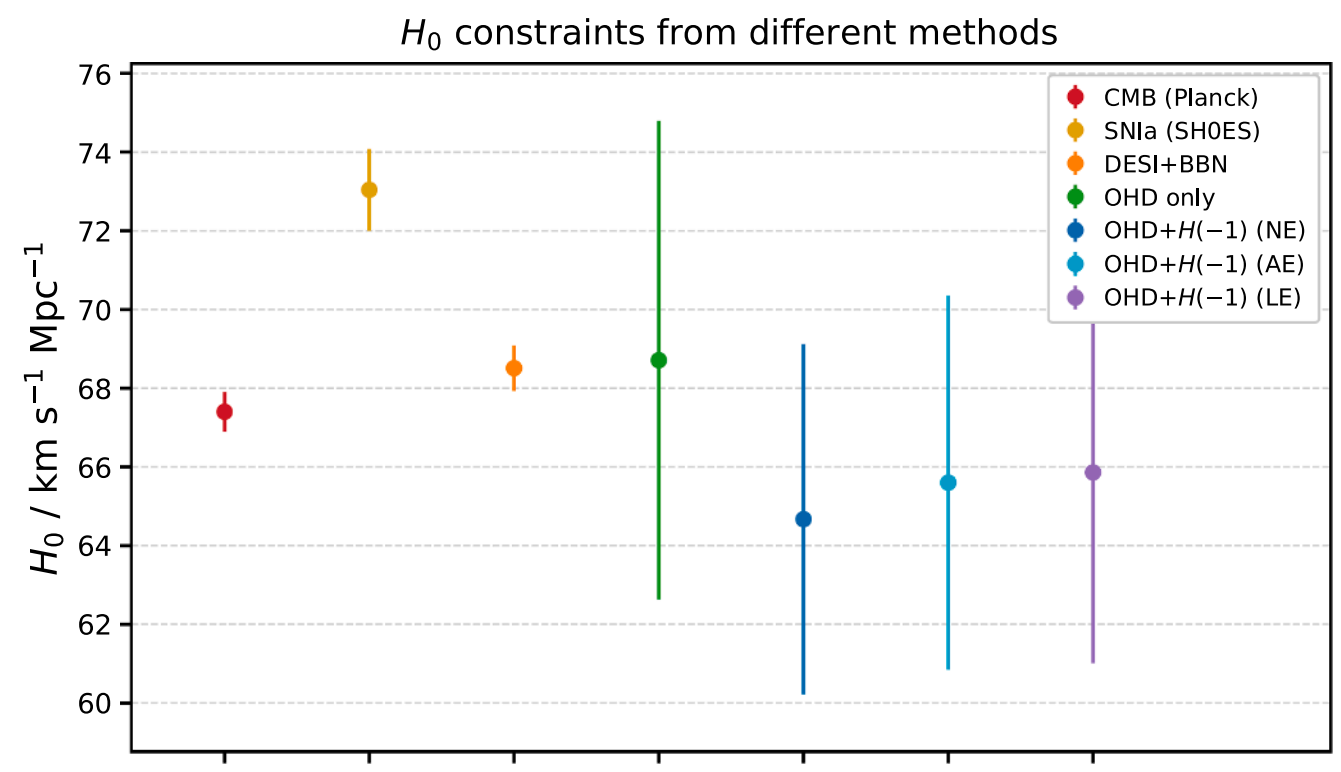

**Fig. 3.** Comparison of $H_0$ constraints from different methods. Shown are the mean values and $1\sigma$ uncertainties for a representative CMB inference, a local distance-ladder (SN Ia) determination, a DESI BAO constraint calibrated by external information (see text), and our GP reconstructions from the CC OHD sample with and without the $H(-1)$ pseudo-point. Incorporating the $H(-1)$ prior shifts the GP-inferred $H_0$ to lower values.

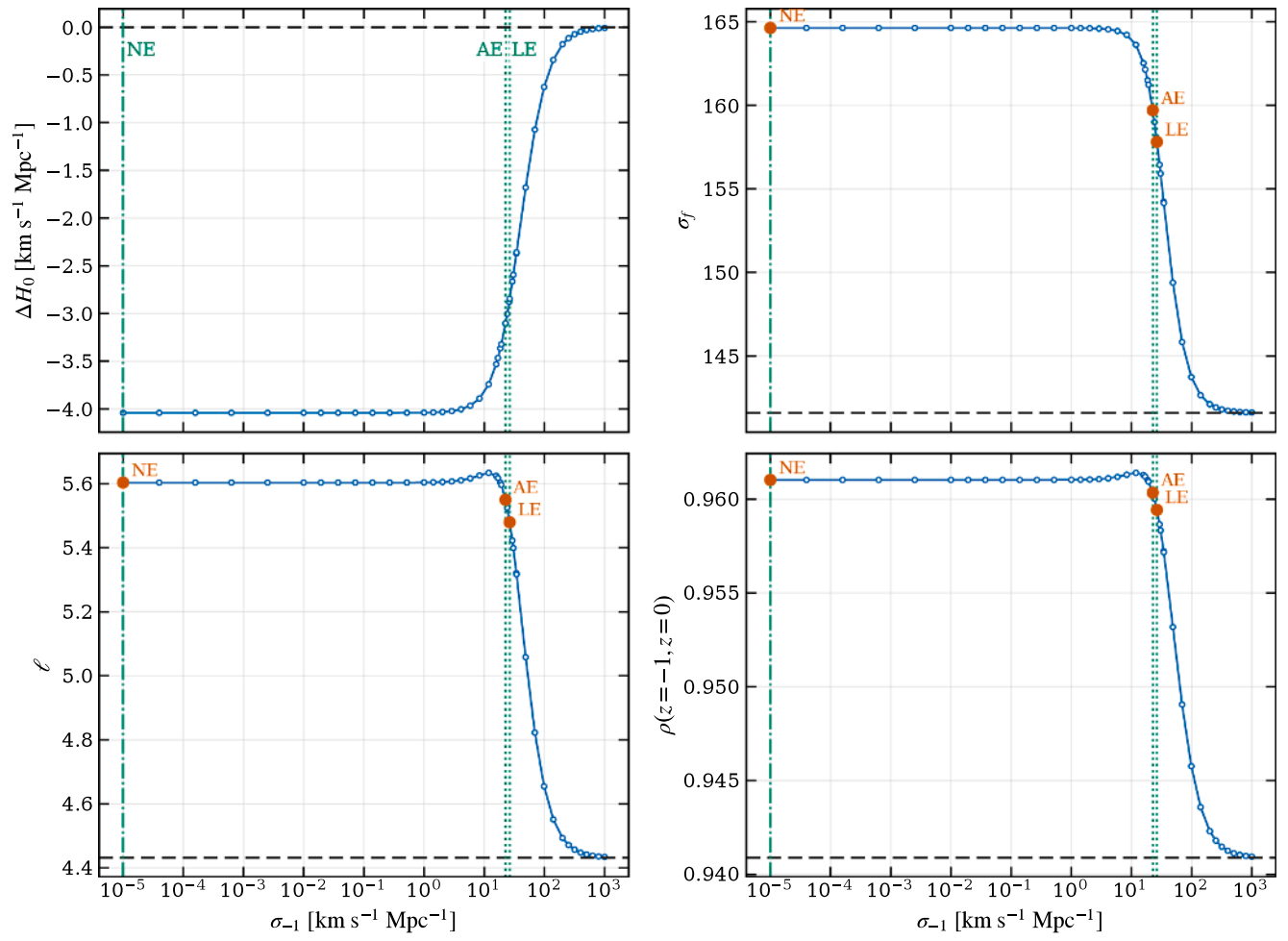

**Fig. 4.** GP-based $\sigma_{-1}$ scan. As the pseudo-point uncertainty is varied, we show the corresponding changes in $\Delta H_0$, the best-fit hyperparameters $\sigma_f$ and $\ell$, and the normalized kernel correlation $\rho(z=-1, z=0)$. Vertical markers indicate the NE/AE/LE choices.

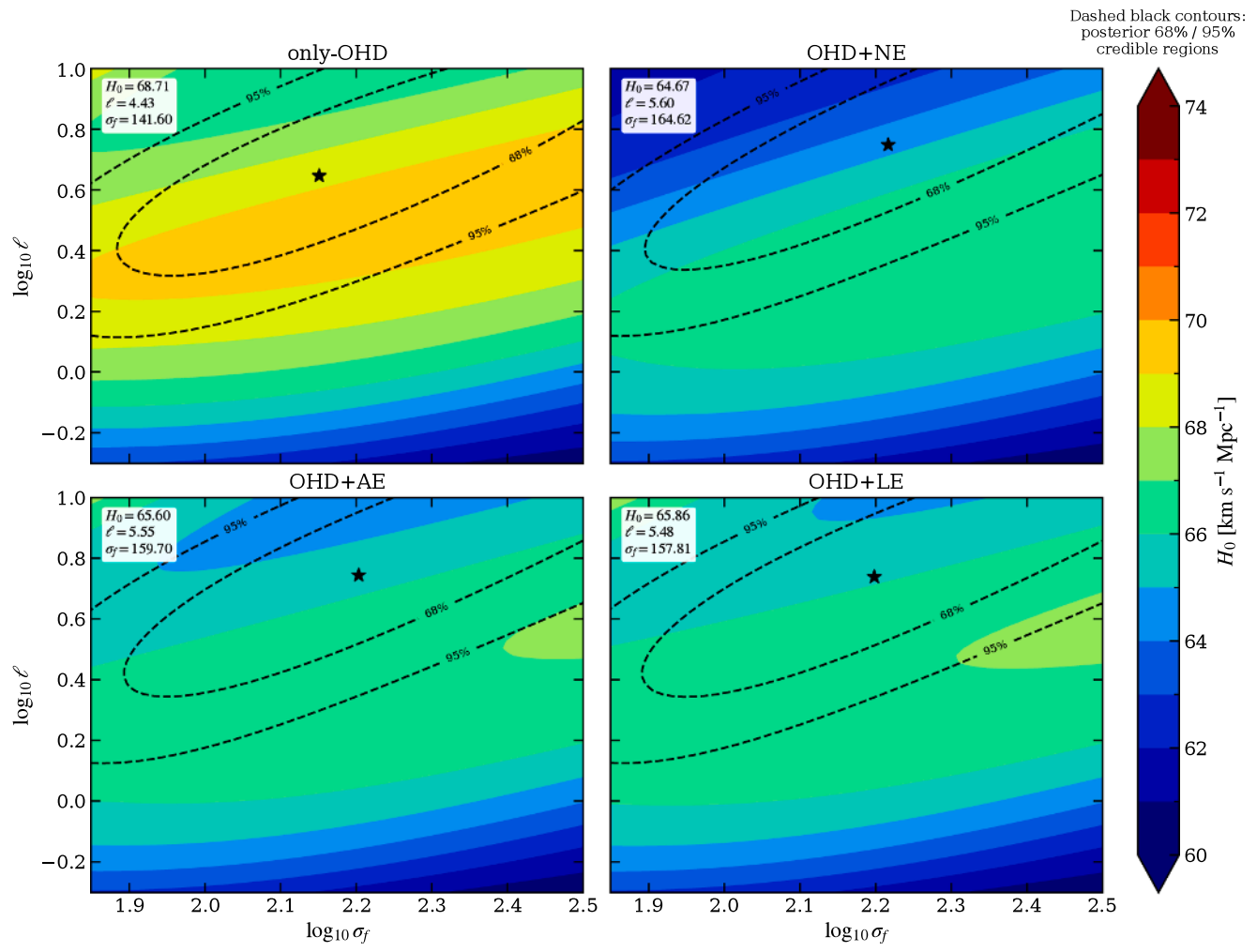

**Fig. 5.** Hyperparameter posterior maps in the $(\log_{10}\sigma_f, \log_{10}\ell)$ plane for the only-OHD, OHD + NE, OHD + AE, and OHD + LE cases. Colors show the conditional GP prediction for $H_0$ at each grid point; dashed contours mark the 68% and 95% highest-posterior-density regions; stars indicate the best-fit hyperparameters.

For the Matérn($\nu = 3/2$) kernel adopted here, this gives

$$\rho(z=-1, z=0) = \left(1 + \frac{\sqrt{3}}{\ell}\right) \exp\left(-\frac{\sqrt{3}}{\ell}\right), \tag{6}$$

Since the pseudo-point is imposed at $z = -1$ and $H_0$ is evaluated at $z = 0$, we use $\rho(z=-1, z=0)$ as a simple diagnostic of their kernel correlation. Fig. 4 shows the resulting trends. As $\sigma_{-1}$ decreases and the future prior is weighted more strongly, the preferred hyperparameters shift away from the OHD-only values, with both $\ell$ and $\rho(z=-1, z=0)$ increasing. This means that the pseudo-point becomes more strongly correlated with the extrapolated present epoch in the GP prior, and the corresponding $\Delta H_0$ becomes more negative. At large $\sigma_{-1}$, the hyperparameters and $\rho(z=-1, z=0)$ return to the OHD-only limit and $\Delta H_0 \to 0$ by $\sigma_{-1} \sim 10^3$.

To further clarify this mechanism, we construct the hyperparameter posterior in the $(\log_{10}\sigma_f, \log_{10}\ell)$ plane on a two-dimensional grid while keeping the kernel family fixed to Matérn($\nu = 3/2$). Following Ref. [39], we evaluate the GP log-marginal likelihood on a logarithmically uniform two-dimensional grid and compute the corresponding conditional prediction for $H_0$ at $z = 0$ at each grid point. The joint and marginal posteriors are then visualized in a form similar to recent GP-based cosmological hyperparameter analyses [40]. For the main posterior scan, we adopt

$$\sigma_f \in [\sqrt{5 \times 10^3},\ \sqrt{10^5}], \qquad \ell \in [0.5,\ 10],$$

sampled with a $48 \times 56$ logarithmically uniform grid. We adopt priors that are flat in $(\log_{10}\sigma_f, \log_{10}\ell)$, and construct the hyperparameter posterior on the scanned domain according to

$$p(\log_{10}\sigma_f, \log_{10}\ell \mid \text{data}) \propto \exp\big[\ln p(\text{data} \mid \sigma_f, \ell)\big], \tag{7}$$

We then normalize this posterior numerically over the sampled grid [17]. The corresponding hyperparameter posterior maps are shown in Fig. 5.

Fig. 5 shows that adding the infinite-future prior systematically shifts the preferred region toward larger $\ell$, with the strongest shift in the NE case and very similar results for AE and LE. At the same time, the joint posterior is elongated along a positively correlated direction, indicating a clear degeneracy between $\sigma_f$ and $\ell$: a larger amplitude and a longer correlation length can produce comparably good fits to the present data. More importantly, lower values of $H_0$ are predominantly associated with larger $\ell$, so the posterior shift toward larger $\ell$ directly moves the preferred GP solutions toward lower $H_0$.

To test whether this trend depends on the particular hyperparameter range adopted in the main scan, we also repeat the analysis over a much wider domain,

$$\sigma_f \in [\sqrt{10^3},\ \sqrt{3 \times 10^8}], \qquad \ell \in [0.5,\ 500],$$

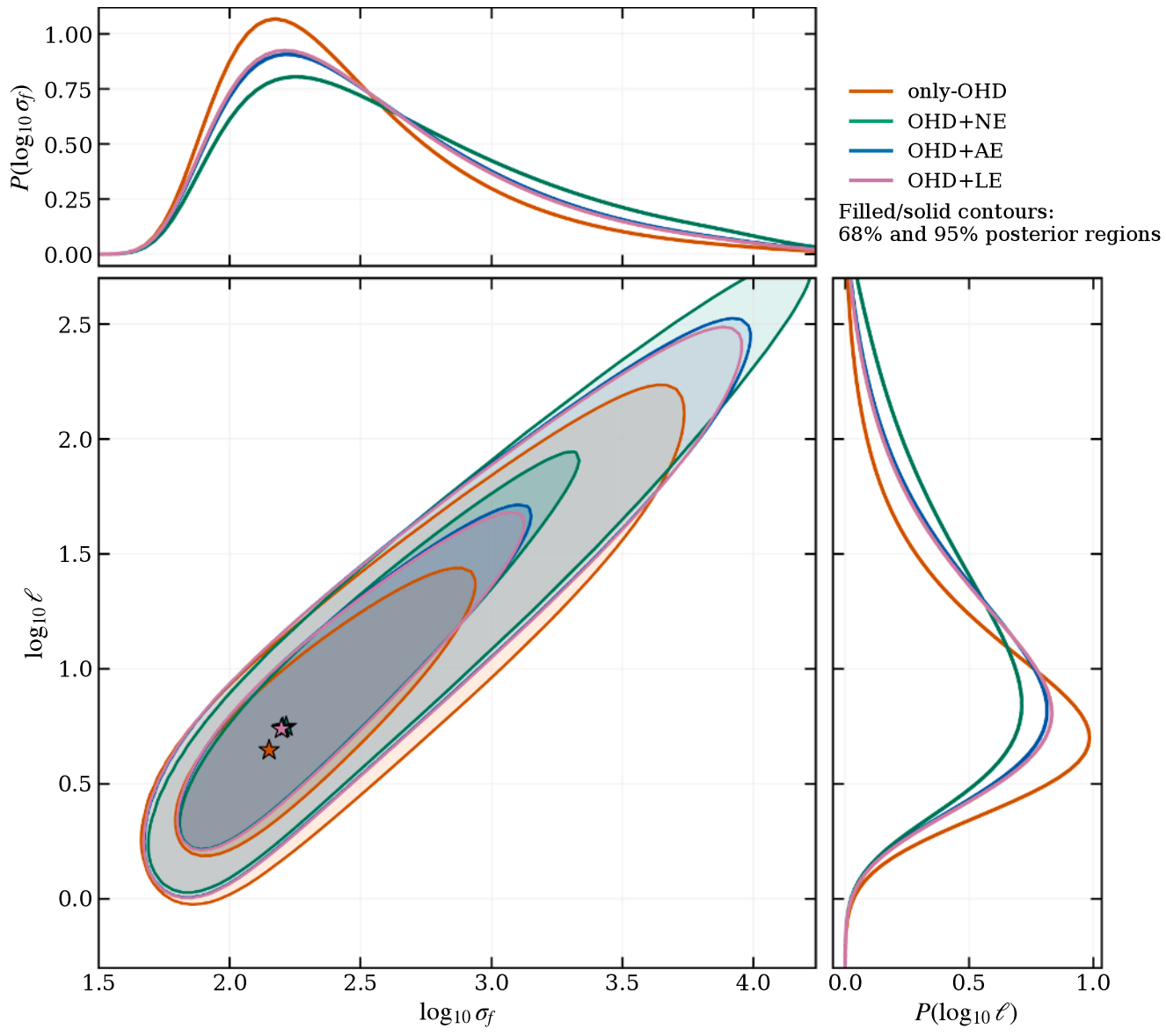

**Fig. 6.** Wide-range hyperparameter posterior maps in the $(\log_{10}\sigma_f, \log_{10}\ell)$ plane. The figure shows the one-dimensional marginal distributions together with the two-dimensional 68% and 95% posterior credible regions for the only-OHD, OHD + NE, OHD + AE, and OHD + LE cases.

using a $78 \times 88$ logarithmically uniform grid. The resulting wide-range posterior maps are shown in Fig. 6. All four cases remain single-peaked and preserve the same relative ordering: the NE case remains most shifted toward larger $\ell$, AE and LE remain very close to each other, and all future-prior cases lie above the only-OHD case in $\ell$.

Taken together with Fig. 4, these posterior maps show that the effect of the future prior is mediated by the GP hyperparameters. As the pseudo-point is assigned a larger weight (smaller $\sigma_{-1}$), the preferred posterior region shifts toward larger $\ell$, which in turn increases $\rho(z=-1, z=0)$ and strengthens the influence of the future pseudo-point on the extrapolated $H_0$ at $z=0$. The downward shift of $H_0$ therefore arises indirectly through this hyperparameter response, rather than from a direct prior on $H_0$ itself.

We then study how the GP-inferred $H_0$ values (with and without the $H(-1)$ pseudo-point) affect parametric constant-$w$CDM fits to the OHD data. We adopt a uniform prior on $w$ over $(-1, 1)$ and perform MCMC sampling using the affine-invariant ensemble sampler (StretchMove). The NE case is not considered here, since it corresponds to an effectively hard enforcement ($\sigma_{-1} \to 0$) and can let the pseudo-point dominate the likelihood, yielding overly tight and potentially biased constraints. Concretely, for each GP reconstruction case, we fix $H_0$ to the corresponding GP-inferred mean value and fit the $w$CDM expansion history to the OHD likelihood using a $\chi^2$ statistic [41,42]. For the flat $w$CDM model, we constrain $(\Omega_m, w)$ and show the marginalized posteriors in Fig. 7. For the non-flat $w$CDM model, we constrain $(\Omega_m, \Omega_{\rm DE}, w)$ and show the results in Fig. 8. For comparison, we repeat the same analysis by fixing $H_0$ to representative external determinations from the local distance ladder [1], *Planck* $\Lambda$CDM [2], and the DESI DR2 BAO calibration using BBN (DESI + BBN) [38]. The resulting parameter constraints are summarized in Table 2.

Table 2 summarizes the parametric $w$CDM fits to the OHD sample for different fixed-$H_0$ reference values, including representative external determinations (CMB, distance ladder, and DESI + BBN) as well as the GP-inferred $H_0$ values obtained in Section 3. In the flat case, $\Omega_m$ varies only mildly across different fixed-$H_0$ reference values (remaining consistent within uncertainties), whereas the inferred equation-of-state parameter $w$ shows a more noticeable dependence on the adopted $H_0$. For example, fixing $H_0$ to the OHD-only GP estimate ($H_0 \simeq 68.71$) yields $w \simeq -0.86 \pm 0.13$, very similar to the DESI + BBN case ($H_0 \simeq 68.51$) and broadly consistent with the CMB-inferred *Planck* $\Lambda$CDM value ($H_0 \simeq 67.4$). By contrast, adopting the lower GP-inferred $H_0$ values obtained when including the $H(-1)$ pseudo-point shifts the posterior toward less negative values, $w \simeq -0.71 \pm 0.16$ (AE) and $w \simeq -0.74 \pm 0.16$ (LE). This indicates that the main impact of incorporating the $H(-1)$ pseudo-point is propagated through the associated shift in the inferred $H_0$, which in turn affects $w$ more strongly than $\Omega_m$ within the assumed prior range $w \in (-1, 1)$. The corresponding posteriors are shown in Fig. 7.

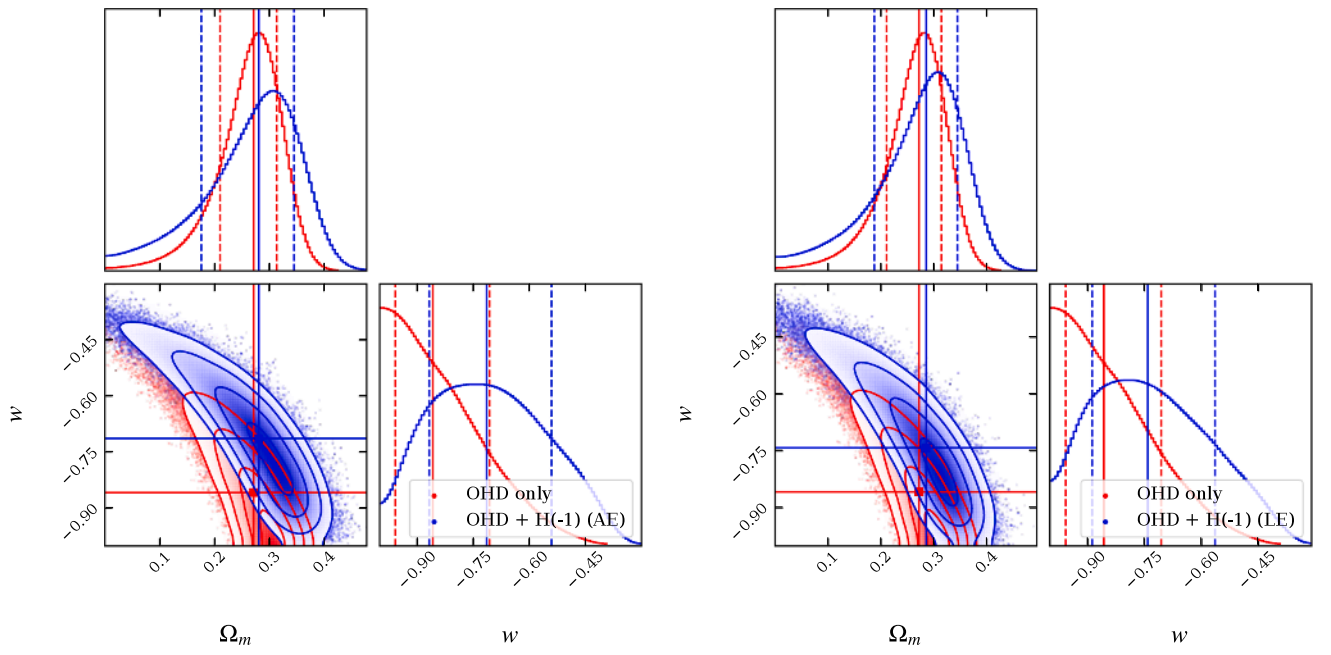

**Fig. 7.** Flat $w$CDM posteriors from OHD-only (red) and OHD + $H(-1)$ (blue) cases. Left: AE; right: LE. Solid (dashed) lines mark the medians (central 68% intervals) of the 1D marginals; shaded contours show the corresponding 2D credible regions.

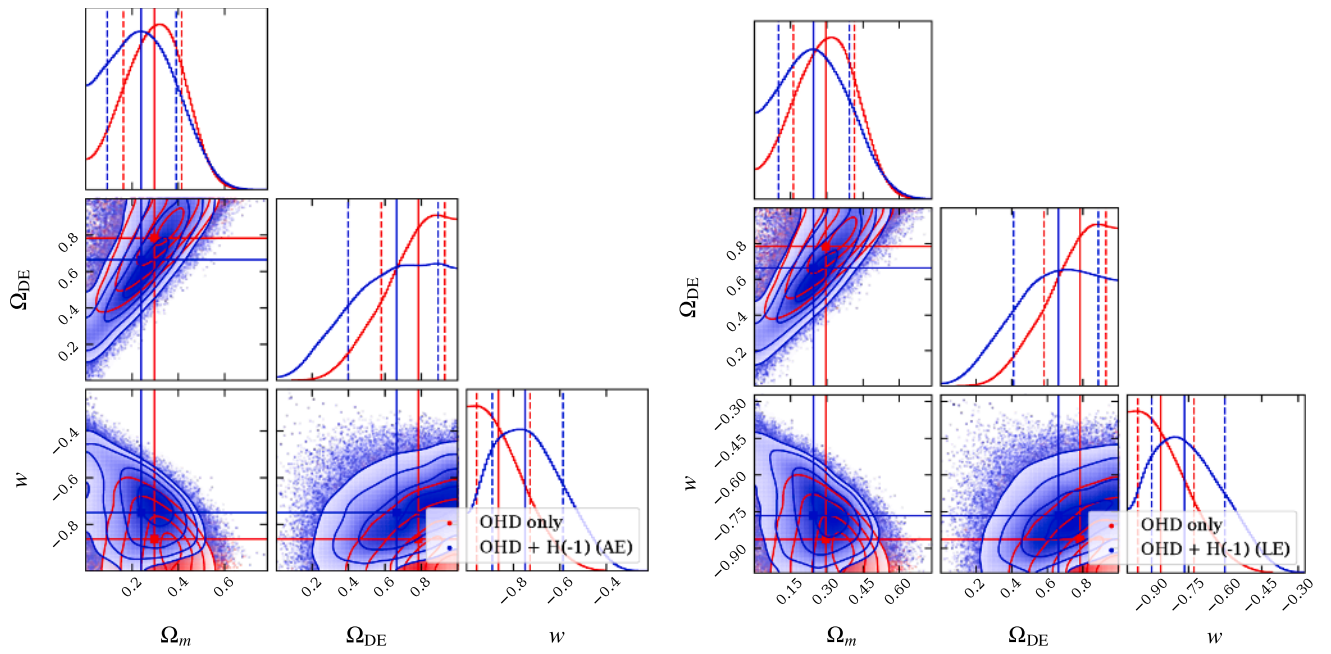

**Fig. 8.** Non-flat $w$CDM posteriors from OHD-only (red) and OHD + $H(-1)$ (blue) cases. Left: AE; right: LE. Line styles and contours follow Fig. 7.

Allowing spatial curvature enlarges the credible regions, reflecting the well-known degeneracies among $\Omega_m$, $\Omega_{\rm DE}$, and $w$ when fitting OHD alone. Consequently, the non-flat constraints in Table 2 are broader than in the flat case. Nevertheless, the same qualitative trend persists: fixing $H_0$ to the lower GP-inferred values obtained with the $H(-1)$ pseudo-point shifts $w$ toward less negative values relative to the OHD-only GP case. At the same time, the posterior medians of $\Omega_m$ and $\Omega_{\rm DE}$ also shift (with $\Omega_m$ moving to smaller values), but these changes are not statistically significant given the enlarged uncertainties. The non-flat posteriors are displayed in Fig. 8. Finally, we assess the curvature preference explicitly using the derived posterior $\Omega_k \equiv 1 - \Omega_m - \Omega_{\rm DE}$ (Fig. 9). While the $\Omega_k$ constraints remain broad and consistent with $\Omega_k = 0$, the cases using the $H(-1)$ GP-inferred $H_0$ reference values mildly shift posterior weight toward $\Omega_k > 0$ (open geometry).

## 4. Mock-data validation

Since the infinite-future condition is imposed through an additional pseudo-point rather than provided by direct observations, it is important to test its impact in controlled situations where the true expansion history is known. We therefore validate the reconstruction pipeline using mock OHD realizations generated from several fiducial constant-$w$ cosmologies, including both quintessence-like cases with $w > -1$ and non-standard cases with $w \le -1$. This allows us to quantify the induced $H_0$

**Table 2**

Constant-$w$CDM constraints from OHD with $H_0$ fixed to representative reference $H_0$ values. For each row, we fix $H_0$ to the value listed in the first column and perform MCMC fits with a uniform prior $w \in (-1, 1)$. Quoted values are marginalized posteriors reported as median $\pm 1\sigma$. External $H_0$ reference values are taken from the local distance ladder [1], *Planck* ΛCDM [2], and DESI DR2 BAO + BBN [38]. The GP-based $H_0$ reference values correspond to the OHD-only reconstruction and the OHD + $H(-1)$ reconstructions under the AE/LE prescriptions.

| $H_0$ [km s$^{-1}$ Mpc$^{-1}$] | $\Omega_m$ (flat) | $w$ (flat) | $\Omega_m$ (non-flat) | $\Omega_{DE}$ (non-flat) | $w$ (non-flat) | $H_0$ reference |
|---|---|---|---|---|---|---|
| 67.40 (fixed) | $0.28 \pm 0.06$ | $-0.82 \pm 0.15$ | $0.28 \pm 0.14$ | $0.74 \pm 0.20$ | $-0.83 \pm 0.14$ | *Planck* |
| 73.04 (fixed) | $0.23 \pm 0.03$ | $-0.93 \pm 0.07$ | $0.31 \pm 0.09$ | $0.89 \pm 0.10$ | $-0.94 \pm 0.06$ | distance ladder |
| 68.51 (fixed) | $0.28 \pm 0.05$ | $-0.86 \pm 0.13$ | $0.29 \pm 0.13$ | $0.77 \pm 0.18$ | $-0.86 \pm 0.12$ | DESI + BBN |
| 68.71 (fixed) | $0.27 \pm 0.05$ | $-0.86 \pm 0.13$ | $0.30 \pm 0.13$ | $0.78 \pm 0.17$ | $-0.86 \pm 0.11$ | OHD only (GP) |
| 65.60 (fixed) | $0.28 \pm 0.08$ | $-0.71 \pm 0.16$ | $0.24 \pm 0.15$ | $0.66 \pm 0.25$ | $-0.75 \pm 0.15$ | OHD + $H(-1)$ (AE) |
| 65.86 (fixed) | $0.29 \pm 0.08$ | $-0.74 \pm 0.16$ | $0.24 \pm 0.15$ | $0.66 \pm 0.24$ | $-0.77 \pm 0.15$ | OHD + $H(-1)$ (LE) |

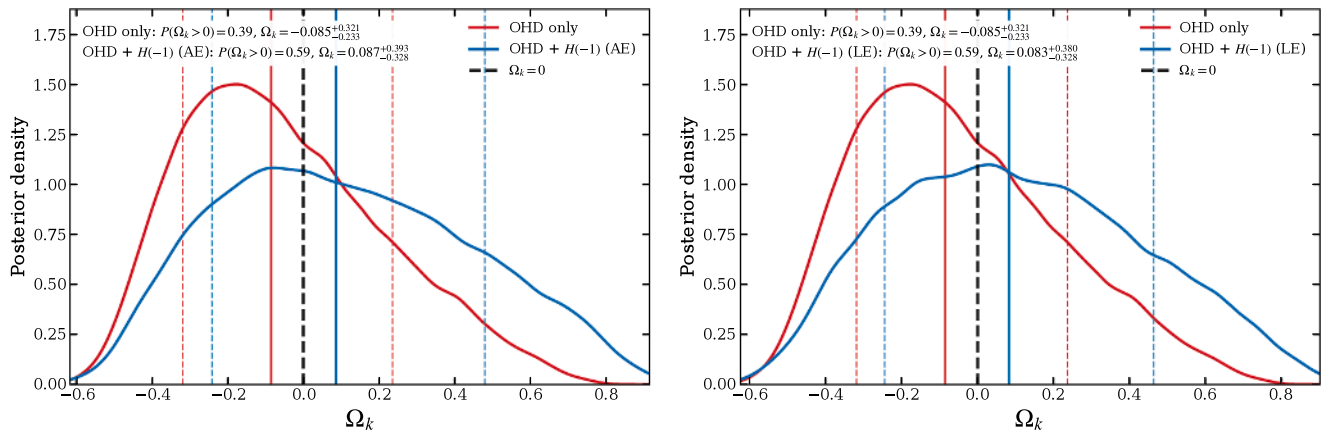

**Fig. 9.** Posterior distributions of $\Omega_k$ derived from the non-flat $\omega$CDM fits. Shown are one-dimensional kernel-density-estimate (KDE) curves for the OHD-only case (red) and the OHD + $H(-1)$ case (blue), using the AE (left) and LE (right) prescriptions for the pseudo-point uncertainty. The vertical dashed black line marks $\Omega_k = 0$. Solid colored vertical lines indicate the posterior medians, and the dashed colored vertical lines indicate the 16th and 84th percentiles. The inset text reports the posterior probability $P(\Omega_k > 0)$ and the median with the central 68% interval.

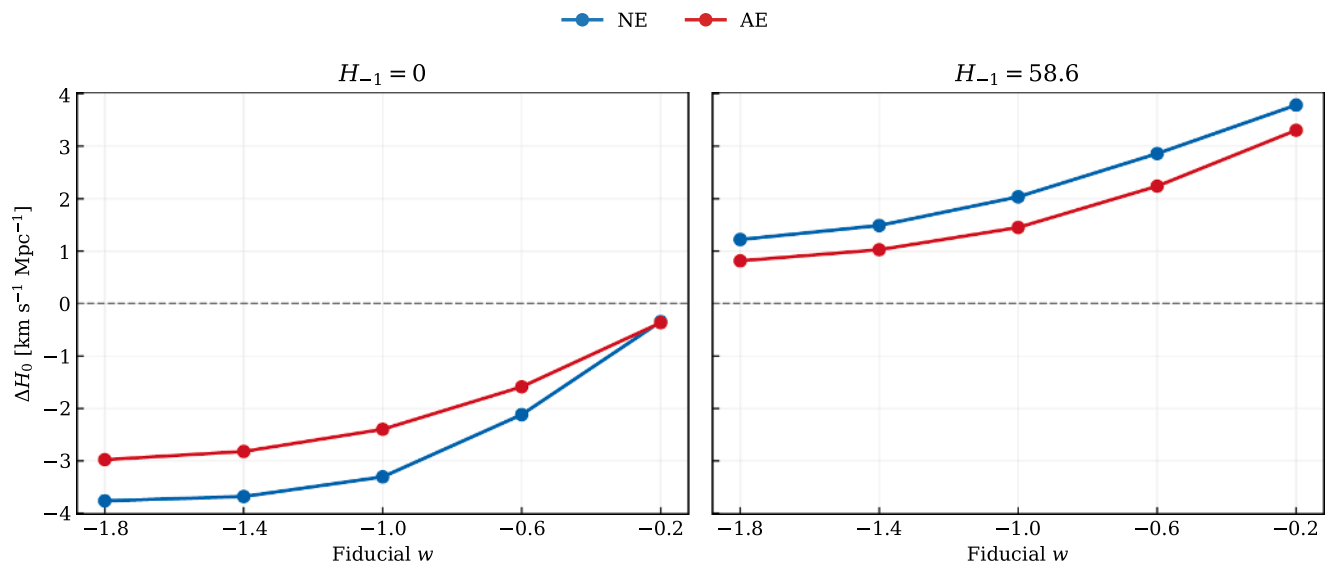

**Fig. 10.** Average shift $\Delta H_0$ relative to the baseline GP reconstruction for mock OHD realizations generated from five fiducial constant-$w$ cosmologies. The left and right panels show the results for the imposed future values $H_{-1} = 0$ and $H_{-1} \simeq 58.6$, respectively. Blue and red markers correspond to the NE and AE prescriptions.

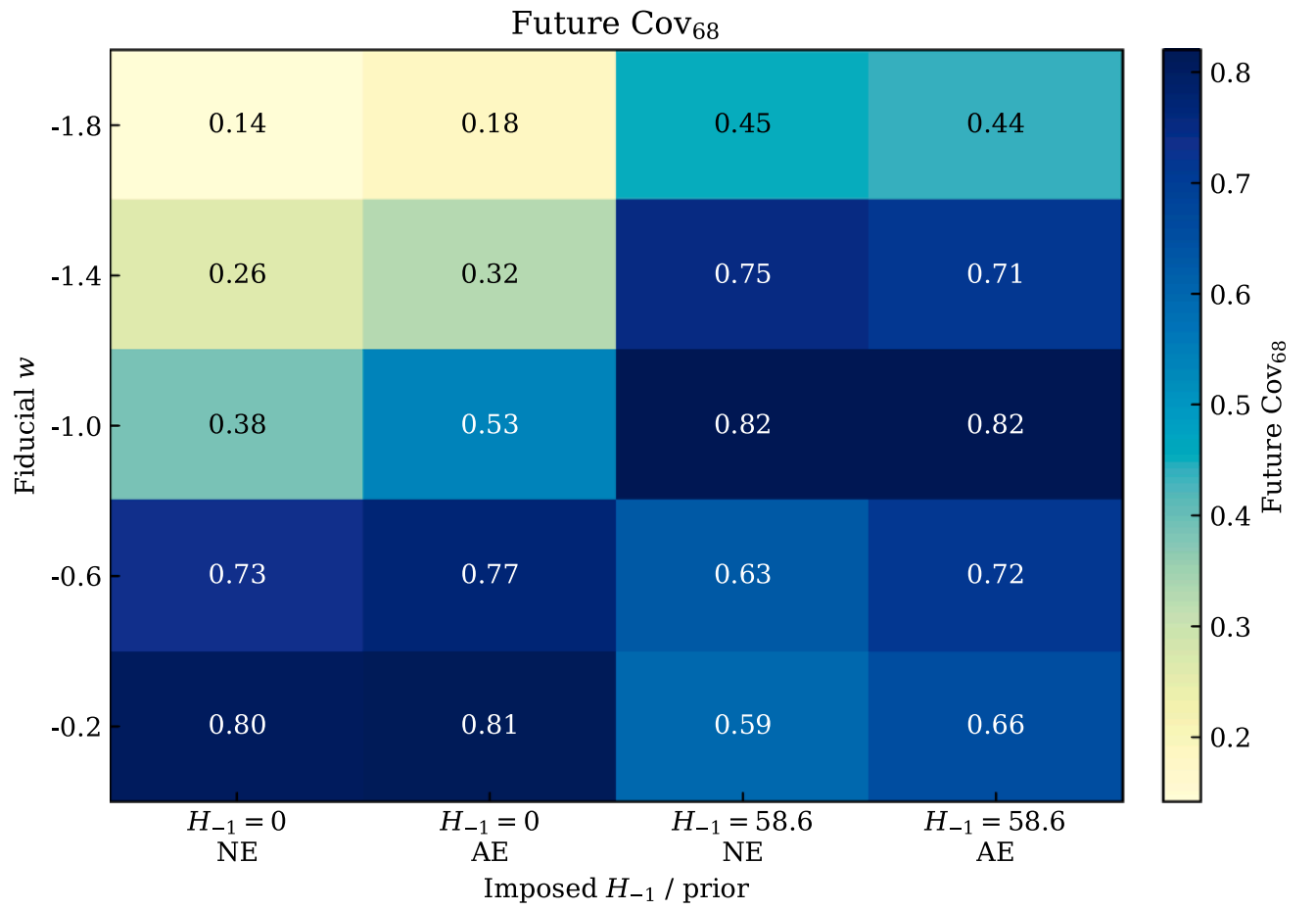

**Fig. 11.** Empirical coverage $\mathrm{Cov}_{68}$ of the reconstructed central 68% interval for $H_0$ for the future-conditioned mock reconstructions. Rows correspond to the five fiducial constant-$w$ cosmologies, while columns show the four combinations of imposed future value and future-point uncertainty considered in this section.

bias and interval coverage when the assumed future condition is either compatible or incompatible with the underlying fiducial cosmology.

For this test, we generate mock OHD samples using the same redshift points and measurement uncertainties as the 37 CC data points employed in the main analysis, while taking the fiducial expansion history to be constant-$w$CDM with $H_0 = 70\ \mathrm{km\,s^{-1}\,Mpc^{-1}}$ and $\Omega_m = 0.3$. Similar mock-$H(z)$ constructions have been considered in Refs. [41,43]. We consider five representative fiducial equations of state,

$$w = -1.8,\ -1.4,\ -1.0,\ -0.6,\ -0.2,$$

and for each fiducial we generate 500 independent mock realizations according to

$$H_i^{\mathrm{mock}} = H_{\mathrm{fid}}(z_i; H_0, \Omega_m, w) + \epsilon_i, \qquad \epsilon_i \sim \mathcal{N}(0, \sigma_{H,i}^2), \tag{8}$$

where $\sigma_{H,i}$ denotes the uncertainty of the corresponding CC measurement. Each mock realization is then analyzed in two ways: first with the baseline GP reconstruction using the mock OHD alone, and then with the same reconstruction after adding an infinite-future pseudo-point. In addition to the main-text choice $H_{-1} = 0$, we also test a control future condition $H_{-1} = H_0\sqrt{1 - \Omega_m} \simeq 58.6$, corresponding to the asymptotic value of the fiducial ΛCDM case with the same $(H_0, \Omega_m)$. For each imposed future value, we consider two representative future-point uncertainties, namely the NE and AE prescriptions defined in Section 2.

To quantify the effect of the imposed future condition, we compute the average shift in $H_0$ relative to the baseline reconstruction,

$$\Delta H_0 \equiv \left\langle \hat{H}_{0,\mathrm{future}} - \hat{H}_{0,\mathrm{base}} \right\rangle, \tag{9}$$

and the empirical coverage of the reconstructed central 68% interval,

$$\mathrm{Cov}_{68} \equiv \frac{1}{N_{\mathrm{mock}}} \sum_{n=1}^{N_{\mathrm{mock}}} \mathbf{1}\Big(H_{0,\mathrm{true}} \in [H_{0,16}^{(n)},\ H_{0,84}^{(n)}]\Big), \tag{10}$$

where $H_{0,\mathrm{true}} = 70\ \mathrm{km\,s^{-1}\,Mpc^{-1}}$ and $[H_{0,16}^{(n)}, H_{0,84}^{(n)}]$ denotes the reconstructed central 68% interval of the $n$th mock realization. In this way, the mock analysis directly quantifies the bias induced by the imposed future condition and evaluates the fraction of realizations in which the reconstructed central 68% interval for $H_0$ contains the true value under the adopted observational error model.

The mock-validation results are summarized in Figs. 10 and 11. Fig. 10 shows the average shift $\Delta H_0$ relative to the baseline reconstruction, while Fig. 11 shows the empirical coverage $\mathrm{Cov}_{68}$ for the future-conditioned reconstructions. For the imposed condition $H_{-1} = 0$, all five fiducial models show a negative $\Delta H_0$, with the largest downward shifts appearing for the phantom-like and ΛCDM-like mocks and a much weaker effect toward the quintessence-like side. When the imposed future value is changed to $H_{-1} \simeq 58.6$, the sign reverses and $\Delta H_0$ becomes positive, with the upward shift growing toward less negative $w$. Taken together, the two panels show that the farther the imposed future-point value departs from the fiducial model prediction for $H(-1)$, the larger the magnitude of the induced shift in $H_0$.

Fig. 11 shows the same trend in the interval coverage. For $H_{-1} = 0$, the future-conditioned reconstructions have low $\mathrm{Cov}_{68}$ for the ΛCDM-like and phantom-like fiducials, with values ranging from 0.14 to 0.53, whereas the quintessence-like cases remain much higher, around 0.73–0.81. For $H_{-1} \simeq 58.6$, the best coverage appears around the ΛCDM-like fiducials, reaching about 0.82, while it decreases again for the quintessence-like models to about 0.59–0.72 and remains poor for the most phantom case, around 0.44. In other words, when the imposed future condition is consistent with the fiducial cosmology's asymptotic value of $H(-1)$, the reconstructed $H_0$ interval contains the true value more often; when the two are incompatible, the coverage becomes noticeably worse.

The NE and AE prescriptions lead to the same qualitative conclusion in both figures. The NE case produces a larger absolute shift in $H_0$ and a stronger change in $\mathrm{Cov}_{68}$ than AE, as expected from the smaller assigned future-point uncertainty. Taken together, these mock tests show that the impact of the future condition on the reconstructed $H_0$ depends on whether the imposed future condition is compatible with the asymptotic future state of the fiducial cosmology.

These mock results also help interpret the real-data results discussed in Section 3. For the observed OHD sample, the baseline reconstruction $H_0 = 68.71 \pm 6.08$ changes to $H_0 = 64.67 \pm 4.45$ after imposing the NE future point with $H_{-1} = 0$, corresponding to a difference of only $0.54\sigma$. In the mock tests, a stronger mismatch between the imposed future condition and the fiducial cosmology is reflected primarily in a poorer coverage of the reconstructed $H_0$ interval. For the real OHD sample, where no such coverage statistic is directly available, the closest analogue is whether the baseline and future-conditioned $H_0$ estimates remain compatible within their quoted $1\sigma$ ranges. The relatively modest shift seen in the data therefore does not exclude the possibility that the real Universe is compatible with a quintessence-like future condition satisfying $H(-1) = 0$; rather, it indicates that the current OHD data do not yet discriminate strongly among different asymptotic future states.

## 5. Conclusions and discussion

In this work we examined how imposing the infinite-future condition $H(z = -1) = 0$ affects non-parametric reconstructions of the expansion history and the inferred Hubble constant. Implemented as a pseudo-point at $z = -1$ in a Gaussian-process reconstruction trained on cosmic-chronometer $H(z)$ data, this prior leads to a systematically lower GP-inferred $H_0$ than the OHD-only case, shifting it numerically closer to representative CMB-inferred $H_0$ values while increasing the discrepancy with the local distance-ladder determination [1]. This downward shift remains robust when repeating the analysis with several commonly used stationary kernels, indicating that the qualitative trend is not sensitive to the kernel choice. Mean-function tests with ΛCDM and Phenomenologically Emergent Dark Energy (PEDE) baselines show that the strongest-prior NE case remains stable in sign, whereas the AE/LE results become strongly mean dependent. Our explicit hyperparameter analysis further shows that the shift is mediated by the GP response to the pseudo-point: as its effective weight increases, the preferred correlation length $\ell$ becomes larger and the kernel correlation between $z = -1$ and $z = 0$ is strengthened, allowing the imposed future condition to propagate more efficiently to the extrapolated $H_0$. Mock-data validation further shows that both the induced $H_0$ bias and the interval coverage depend on whether the imposed future condition is compatible with the true asymptotic future state of the underlying cosmology. However, given that the GP uncertainty still dominates at present, the downward shift in $H_0$ is more naturally interpreted as the imposed asymptotic boundary condition propagating to low redshift through the GP correlation structure and affecting the statistical extrapolation. It should not be taken as evidence for direct support from early-Universe information or any physical link.

We then performed parametric constant-$w$CDM fits to the OHD sample with $H_0$ fixed to several representative reference values (Table 2). In the flat case, $\Omega_m$ is only weakly affected, whereas $w$ shifts toward less negative values when adopting the lower $H_0$ reference values. Allowing spatial curvature enlarges the credible regions due to the known degeneracies among $\Omega_m$, $\Omega_{\mathrm{DE}}$, and $w$ when fitting OHD alone; nevertheless, the same qualitative trend with the fixed-$H_0$ reference values persists. The derived curvature posterior remains broad and consistent with $\Omega_k = 0$, with the cases using the $H(-1)$ GP-inferred $H_0$ reference values mildly shifting posterior weight toward $\Omega_k > 0$ (open geometry).

In interpreting these results, it is useful to separate the boundary-value construction from its model motivation. In an FLRW background, $1 + z = a_0/a(t)$ maps the infinite-future limit $a \to \infty$ to $z \to -1$, so $z < 0$ is only a formal extension of the redshift variable, not an observationally accessible domain. The condition $H(-1) = 0$ should therefore be interpreted strictly as a theoretical boundary prior. In the GP reconstruction this prior is compressed into a single pseudo-point at $z = -1$ with tunable uncertainty $\sigma_{-1}$. Scanning $\sigma_{-1} \in [10^{-5}, 10^3]$ shows that the $H_0$ shift interpolates between the strong-prior regime and the data-dominated OHD-only limit, demonstrating that the effect is controlled by the pseudo-point weight and transmitted through the GP correlation structure.

This distinction is important when interpreting the method in light of recent DESI results. When DESI BAO measurements are combined with CMB and Type Ia supernova information, several analyses report a preference for dynamical dark energy whose statistical significance depends on the dataset combination and supernova compilation, reaching the $\sim 3\sigma$ level or above in some cases [38,44]. In our GP reconstruction framework, the future limiting value of $H(z)$ depends on the specific dark-energy model assumption. The GP itself cannot identify which cosmological model produced the imposed pseudo-point; it only receives the boundary value $H(-1) = 0$ with an assigned weight. Hence the practical role of the present pseudo-point construction with current data is to quantify how an assumed, model-dependent future boundary value propagates through the GP covariance structure and affects the inferred $H_0$, rather than to provide a model-selection test among full dark-energy dynamical models.

Such a construction can be extended beyond constant-$w$CDM only when the underlying model or parametrization supplies a physically meaningful future boundary condition. For example, if the Chevallier–Polarski–Linder form [45,46] is assumed to remain valid into the asymptotic future, the restricted prior $w_a < 0$ (or the constant-$w$ limit $w_a = 0$ with $w_0 > -1$) formally gives $H(z \to -1) = 0$ in a flat or otherwise non-recollapsing background. In models with a negative or sign-switching cosmological constant [47,48], the future evolution can instead have a finite lifetime and end at a turning point where $H = 0$, rather than approach a simple $z \to -1$ asymptote. Moreover, a single pseudo-point encodes only the asymptotic value and cannot distinguish models that share the same $H(-1)$ but approach it differently. Capturing such differences would require additional model-dependent information, such as future pseudo-points at finite negative redshifts, derivative constraints near $z = -1$, or a dynamical model imposed at the level of the expansion equations. Related derivative-based anchors have been discussed in recent work [49], but implementing such constraints would require assumptions beyond the present GP framework applied directly to $H(z)$. We therefore leave these extensions for future work.

## Data availability

Data will be made available on request.

## Declaration of competing interest

The authors declare the following financial interests/personal relationships which may be considered as potential competing interests: Tong-Jie Zhang reports financial support was provided by National SKA

Program of China. Tong-Jie Zhang reports financial support was provided by China Manned Space Program (CMS-CSST). Cheqiu Lyu reports financial support was provided by Fundamental Research Funds for the Central Universities. If there are other authors, they declare that they have no known competing financial interests or personal relationships that could have appeared to influence the work reported in this paper.

## Acknowledgements

We thank the referee for constructive comments that helped improve this manuscript. This work was supported by the National SKA Program of China (No. 2022SKA0110202) and the China Manned Space Program (CMS-CSST-2025-A01); CL is supported by the Fundamental Research Funds for the Central Universities (WK2030250123).

## Appendix A. GP robustness tests

### A.1. Kernel robustness

Motivated by recent Bayesian kernel studies emphasizing that kernel screening should be guided by the physical quantity of interest [26,50], we perform a dedicated kernel-robustness test focused on the pseudo-point induced $H_0$ shift. Specifically, we repeat the GP reconstruction using the same OHD sample as in the main text and, separately, using the OHD sample augmented by $H(-1) = 0$ under the NE/AE/LE prescriptions defined in Section 2. In all cases, we keep the GP setup identical to the main analysis and vary only the kernel function. The kernel hyperparameters are determined by maximizing the GP log-marginal likelihood (LML) [17].

In addition to the Matérn kernel adopted in the main text, we consider several widely used stationary kernels for cosmological reconstructions. For completeness, we list here the squared-exponential (SE) kernel, also known as the radial-basis-function (RBF) kernel, and the rational-quadratic (RQ) kernel:

$$k_{\mathrm{SE}}(z, z') = \sigma_f^2 \exp\left[-\frac{(z - z')^2}{2\ell^2}\right], \tag{A.1}$$

$$k_{\mathrm{RQ}}(z, z') = \sigma_f^2 \left(1 + \frac{(z - z')^2}{2\alpha\ell^2}\right)^{-\alpha}, \tag{A.2}$$

where $\sigma_f$ and $\ell$ denote the characteristic amplitude and correlation length, and $\alpha$ is an additional shape parameter in the RQ kernel. (The Matérn kernel definition is given in Eq. (3) of the main text.)

Fig. A.1 summarizes the inferred $H_0$ constraints obtained from different kernels for the OHD-only reconstruction and for the reconstructions including the $H(-1) = 0$ pseudo-point (NE/AE/LE). We find that the qualitative impact of imposing $H(-1) = 0$ is stable against the kernel choice: across all tested kernels, the inclusion of the pseudo-point shifts the reconstructed $H_0$ toward lower values relative to the OHD-only case.

Table A.1 lists the corresponding numerical results and the optimized LML values for each kernel and case. We find that the optimized LML values are broadly comparable among the tested kernels within a given case, indicating that no single kernel is strongly preferred by the data. In particular, the Matérn($\nu = 3/2$) kernel adopted in the main text yields an LML comparable to the best-performing alternatives. More importantly, the qualitative impact of imposing the pseudo-point is robust: for all tested kernels, adding $H(-1) = 0$ (NE/AE/LE) shifts the reconstructed $H_0$ to lower values relative to the OHD-only reconstruction.

### A.2. Mean-function robustness

To test the sensitivity of the reconstruction to the GP mean function, we replace the default zero mean by two cosmologically motivated choices based on ΛCDM and PEDE models. Similar mean-function tests have recently been discussed in Refs. [25,51]. In both cases we write

$$H(z) = m(z; \phi) + \delta H(z), \tag{A.3}$$

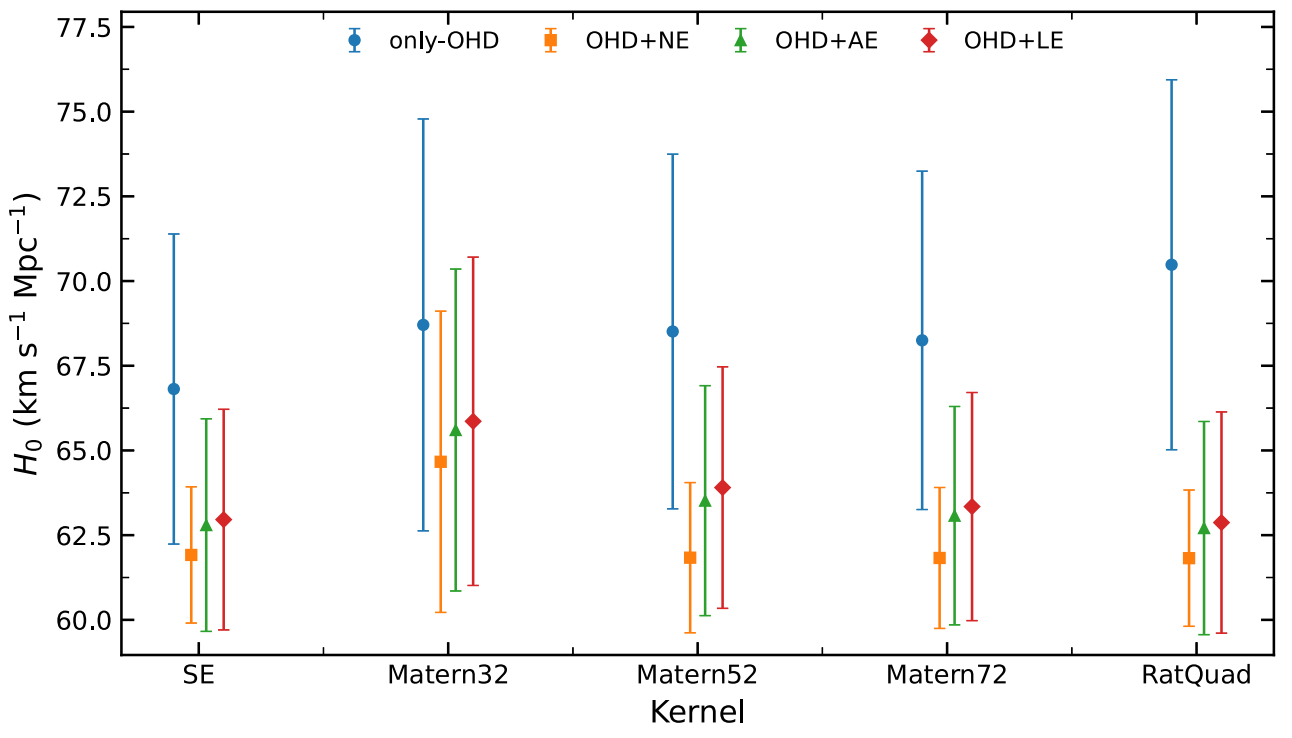

**Fig. A.1.** Kernel robustness of the GP-inferred $H_0$. Shown are the mean values and $1\sigma$ uncertainties of $H_0$ obtained from GP reconstructions using different stationary kernels, for the OHD-only case and for the OHD sample augmented by the pseudo-point $H(-1) = 0$ under the NE/AE/LE prescriptions.

**Table A.1**
Summary of the GP-inferred $H_0$ constraints and the optimized log-marginal likelihood (LML) for different kernels, for the OHD-only reconstruction and for the reconstructions including the pseudo-point $H(-1) = 0$ under the NE/AE/LE prescriptions.

| Kernel | Case | $H_0$ [km s$^{-1}$ Mpc$^{-1}$] | LML |
|---|---|---|---|
| SE | only-OHD | $66.813 \pm 4.576$ | −158.14 |
| Matérn$_{3/2}$ | only-OHD | $68.706 \pm 6.078$ | −157.88 |
| Matérn$_{5/2}$ | only-OHD | $68.511 \pm 5.234$ | −157.89 |
| Matérn$_{7/2}$ | only-OHD | $68.250 \pm 4.993$ | −157.97 |
| RQ | only-OHD | $70.481 \pm 5.460$ | −157.82 |
| SE | OHD + NE | $61.916 \pm 2.011$ | −162.33 |
| Matérn$_{3/2}$ | OHD + NE | $64.665 \pm 4.446$ | −162.99 |
| Matérn$_{5/2}$ | OHD + NE | $61.835 \pm 2.216$ | −162.78 |
| Matérn$_{7/2}$ | OHD + NE | $61.827 \pm 2.080$ | −162.59 |
| RQ | OHD + NE | $61.821 \pm 2.011$ | −162.33 |
| SE | OHD + AE | $62.797 \pm 3.138$ | −162.78 |
| Matérn$_{3/2}$ | OHD + AE | $65.603 \pm 4.751$ | −163.00 |
| Matérn$_{5/2}$ | OHD + AE | $63.519 \pm 3.394$ | −162.96 |
| Matérn$_{7/2}$ | OHD + AE | $63.075 \pm 3.223$ | −162.90 |
| RQ | OHD + AE | $62.709 \pm 3.146$ | −162.79 |
| SE | OHD + LE | $62.959 \pm 3.255$ | −162.87 |
| Matérn$_{3/2}$ | OHD + LE | $65.861 \pm 4.845$ | −163.01 |
| Matérn$_{5/2}$ | OHD + LE | $63.904 \pm 3.565$ | −163.00 |
| Matérn$_{7/2}$ | OHD + LE | $63.344 \pm 3.366$ | −162.96 |
| RQ | OHD + LE | $62.872 \pm 3.265$ | −162.87 |

where $m(z; \phi)$ denotes the parametric mean function, while $\delta H(z)$ describes the GP residual around it. For the two representative mean functions considered here, we take the ΛCDM mean function and the PEDE mean function introduced in Refs. [52,53]:

$$m_{\Lambda\mathrm{CDM}}(z; \phi_\Lambda) = H_0 \sqrt{\Omega_m (1+z)^3 + (1 - \Omega_m)}, \quad \phi_\Lambda = (H_0, \Omega_m), \tag{A.4}$$

$$m_{\mathrm{PEDE}}(z; \phi_{\mathrm{P}}) = H_0 \left[ \Omega_m (1+z)^3 + (1 - \Omega_m)\left[1 - \tanh\left(\log_{10}(1+z)\right)\right] \right]^{1/2}, \quad \phi_{\mathrm{P}} = (H_0, \Omega_m). \tag{A.5}$$

For each mean-function choice, we first constrain the mean-function parameters $\phi$ using the only-OHD sample, and draw posterior samples $\phi^{(s)}$ with MCMC.

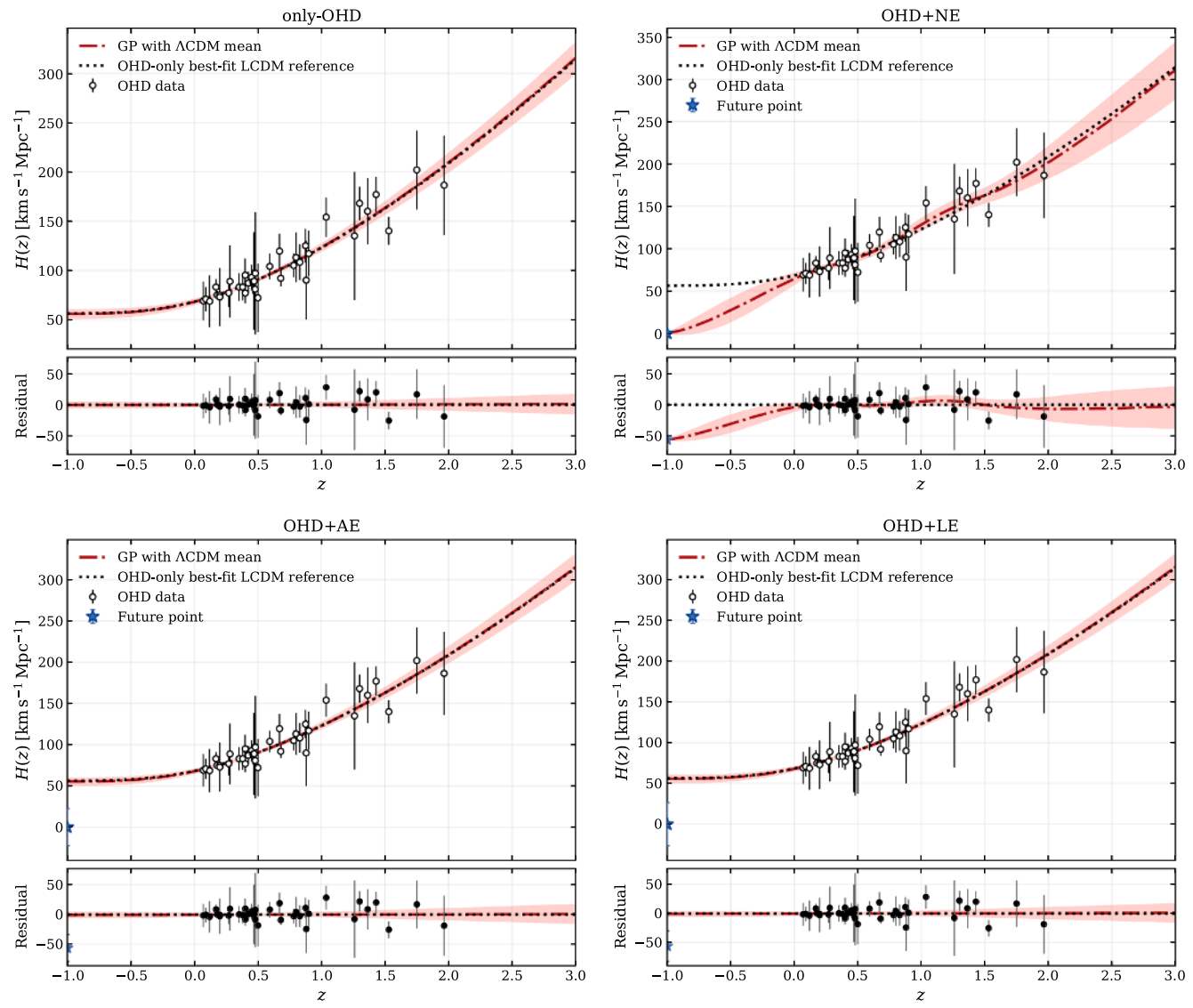

**Fig. A.2.** GP reconstructions obtained with the ΛCDM mean function. The four sub-panels correspond to the only-OHD, OHD + NE, OHD + AE, and OHD + LE cases, respectively. Solid curves show the median reconstruction, and shaded bands denote the central 68% credible intervals.

For each sampled mean-function parameter set $\phi^{(s)}$, we repeat the residual-GP step for the four cases

$C \in \{\text{only-OHD}, \text{OHD+NE}, \text{OHD+AE}, \text{OHD+LE}\}$.

On the corresponding training set we construct residuals

$$r_i^{(s,C)} = H_i^{(C)} - m\left(z_i^{(C)}; \phi^{(s)}\right), \tag{A.6}$$

where the future pseudo-point is included only at this residual-GP stage for the OHD + NE/AE/LE cases. We model the residual field as

$$\delta H^{(s,C)}(z) \sim \mathcal{GP}\big(0,\ k_{\theta(s,C)}(z, z')\big), \tag{A.7}$$

so that the zero-mean reconstruction used in the main text is recovered as the special case $m(z; \phi) = 0$, for which $H(z) = \delta H(z)$. We again adopt the same Matérn($\nu = 3/2$) kernel as in the main text. For each $(s, C)$ pair, the residual-GP hyperparameters $\theta^{(s,C)} = (\sigma_f, \ell)$ are re-optimized by maximizing the GP log-marginal likelihood on that residual training set.

After fitting the residual GP, we evaluate its posterior mean $\mu_{\rm GP}^{(s,C)}(z)$ and covariance $\Sigma_{\rm GP}^{(s,C)}(z, z')$ on the reconstruction grid, and generate GP posterior draws

$$\delta H_{\rm draw}^{(s,C)}(z) \sim \mathcal{N}\left(\mu_{\rm GP}^{(s,C)}(z),\ \Sigma_{\rm GP}^{(s,C)}(z, z')\right). \tag{A.8}$$

The full reconstructed history for a generic draw is then

$$H_{\rm draw}^{(s,C)}(z) = m\big(z; \phi^{(s)}\big) + \delta H_{\rm draw}^{(s,C)}(z). \tag{A.9}$$

For each case, we construct an ensemble of reconstructed expansion histories by combining 1000 posterior samples of the mean-function parameters with 10 conditional GP realizations per sample, which we find to provide stable reconstructions and uncertainty bands. The reconstructed $H(z)$ curve and its uncertainty band are then summarized by the median and central 68% credible interval of this ensemble at each redshift, while the corresponding $H_0$ constraint is obtained from the same ensemble evaluated at $z = 0$.

The resulting reconstructions for the ΛCDM and PEDE mean functions are shown in Figs. A.2 and A.3, and the corresponding $H_0$ summaries are listed in Table A.2.

The results show that all three future-prior cases still shift $H_0$ downward relative to the corresponding only-OHD baseline when a nonzero mean function is adopted, but the magnitude of the shift depends strongly on the mean choice. For the strongest-prior case OHD + NE, the reduction remains sizable, about 3.6 km s$^{-1}$ Mpc$^{-1}$ for the ΛCDM mean and about 6.0 km s$^{-1}$ Mpc$^{-1}$ for the PEDE mean. For the milder AE and LE prescriptions, however, the shifts shrink to only ∼ 0.4–0.9 km s$^{-1}$ Mpc$^{-1}$, much smaller than the corresponding zero-mean shifts of about 3.1 and 2.9 km s$^{-1}$ Mpc$^{-1}$. These appendix tests therefore indicate that the downward NE effect is robust in sign across mean-function choices, whereas the AE/LE results are not quantitatively robust once a nonzero mean is introduced.

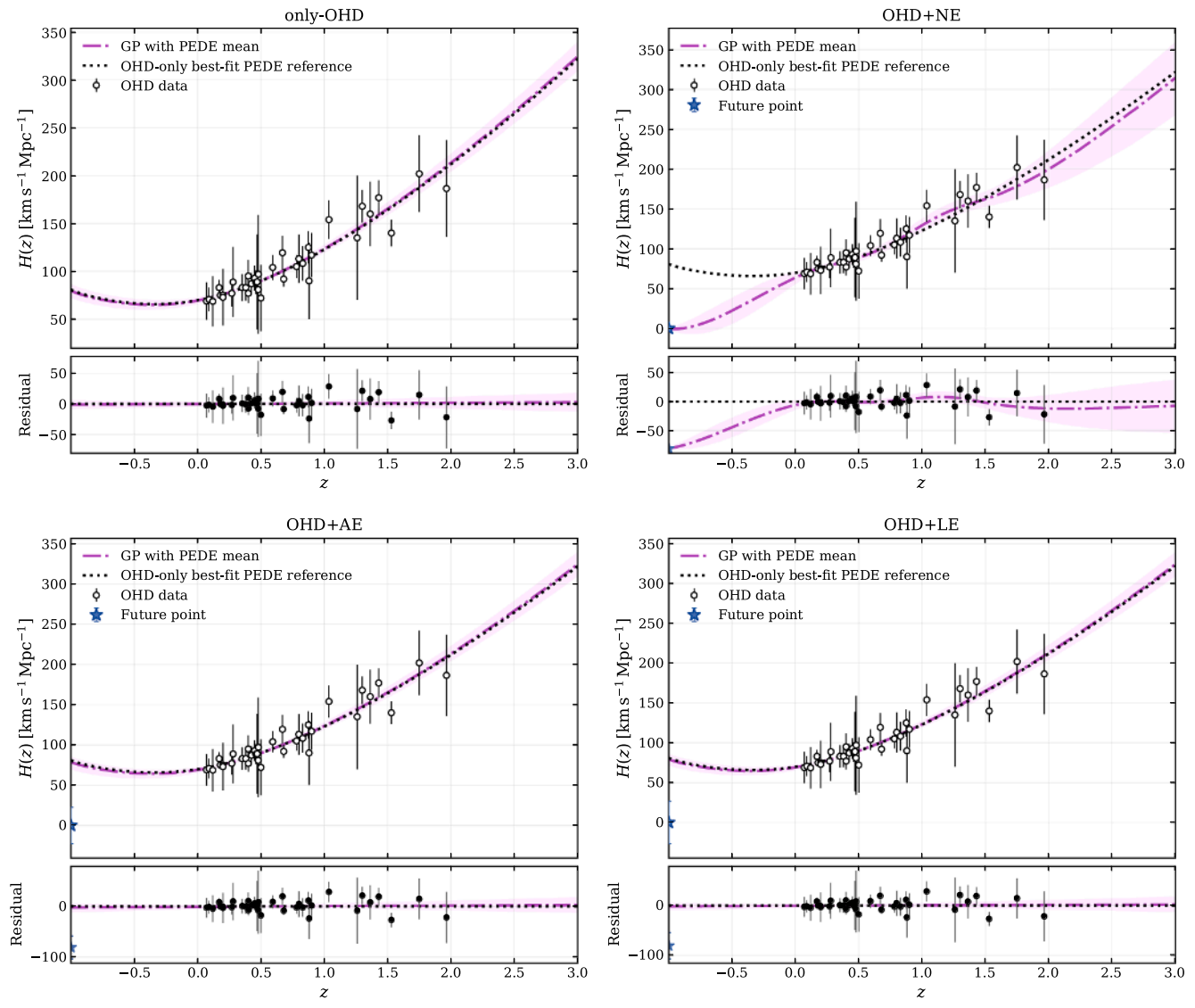

**Fig. A.3.** GP reconstructions obtained with the PEDE mean function. The four sub-panels correspond to the only-OHD, OHD + NE, OHD + AE, and OHD + LE cases, respectively. Solid curves show the median reconstruction, and shaded bands denote the central 68% credible intervals.

**Table A.2**
GP-inferred $H_0$ constraints for different mean-function choices.

| Mean function | Case | $H_0$ [km s$^{-1}$ Mpc$^{-1}$] |
|---|---|---|
| Zero mean | only-OHD | $68.71 \pm 6.08$ |
| | OHD + NE | $64.67 \pm 4.45$ |
| | OHD + AE | $65.60 \pm 4.75$ |
| | OHD + LE | $65.86 \pm 4.85$ |
| ΛCDM mean | only-OHD | $67.98^{+2.50}_{-2.50}$ |
| | OHD + NE | $64.38^{+6.56}_{-6.45}$ |
| | OHD + AE | $67.76^{+2.46}_{-2.38}$ |
| | OHD + LE | $67.87^{+2.46}_{-2.48}$ |
| PEDE mean | only-OHD | $69.44^{+2.71}_{-2.74}$ |
| | OHD + NE | $63.81^{+7.29}_{-7.20}$ |
| | OHD + AE | $68.95^{+2.65}_{-2.71}$ |
| | OHD + LE | $69.16^{+2.67}_{-2.70}$ |